\documentstyle[aps,prl,twocolumn,epsfig]{revtex}

\newcommand{\be}{\begin{equation}}
\newcommand{\ee}{\end{equation}}
\newcommand{\ba}{\begin{eqnarray}}
\newcommand{\ea}{\end{eqnarray}}

\draft
\begin{document}

\twocolumn[\hsize\textwidth\columnwidth\hsize\csname @twocolumnfalse\endcsname

\title{\bf How to Measure Subdiffusion Parameters}

\author{T. Koszto{\l}owicz$^1$, K. Dworecki$^1$, and
St. Mr\'owczy\'nski$^{1,2}$}

\address{$^1$Institute of Physics, \'Swi\c etokrzyska Academy,
ul. \'Swi\c etokrzyska 15, PL - 25-406 Kielce, Poland \\
$^2$So\l tan Institute for Nuclear Studies,
ul. Ho\.za 69, PL - 00-681 Warsaw, Poland}

\date{11-th April 2005}

\maketitle

\vspace{-0.5cm}

\begin{abstract}

We propose a method to measure the subdiffusion parameter $\alpha$ 
and subdiffusion coefficient $D_{\alpha}$ which are defined 
by means of the relation $\langle x^2\rangle =\frac{2D_\alpha}
{\Gamma( 1+\alpha)}\,t^\alpha$ where $\langle x^2\rangle$ denotes 
a mean square displacement of a random walker starting from $x=0$ 
at the initial time $t=0$. The method exploits a membrane system 
where a substance of interest is transported in a solvent from one 
vessel to another across a thin membrane which plays here only an 
auxiliary role. We experimentally study a diffusion of glucose and 
sucrose in a gel solvent, and we precisely determine the 
parameters $\alpha$ and $D_{\alpha}$, using a fully analytic 
solution of the fractional subdiffusion equation.

\end{abstract}

]
\begin{narrowtext}

Subdiffusion occurs in various systems. We mention  here 
a diffusion in porous media or charge carriers transport 
in amorphous semiconductors \cite{mk,bg}. The subdiffusion is 
characterized by a time dependence of the mean square displacement 
of a Brownian particle. When the particle starts form $x=0$ at the 
initial time $t=0$ this dependence in a one-dimension system is 
\be
\label{a}
\left< x^2\right> =\frac{2D_\alpha}
{\Gamma\left( 1+\alpha\right)} \: t^\alpha\,,
\ee
where $D_{\alpha}$ is the subdiffusion coefficient measured in 
the units ${\rm [m^2/s^\alpha ]}$ and $\alpha$ obeys $0< \alpha \le 1$. 
For $\alpha =1$ one deals with the normal or Gaussian diffusion
characterized by the linear growth of $\langle x^2\rangle$ with $t$ 
which results from the Central Limit Theorem applied to many 
independent jumps of a random walker. The anomalous diffusion occurs when 
the theorem fails to describe the system because the distributions 
of summed random variables are too broad or the variables are correlated 
to each other. The subdiffusion is related to infinitely long average 
time that a random walker waits to make a finite jump. Then, its 
average displacement squared, which is observed in a finite time 
interval, is suppressed.

The subdiffusion has been recently extensively studied, see 
{\it e.g.} \cite{mk,bg,bmk,km}. While the phenomenon is 
theoretically rather well understood there are very a few 
experimental investigations. There is no effective method to 
experimentally measure $\alpha$ and $D_{\alpha}$. In the 
pioneering study \cite{km}, where $D_{\alpha}$was determined 
experimentally for the first time, the interdiffusion of heavy 
and light water in a porous medium was observed by means 
of NMR. $D_\alpha$ was found, using the special case $\alpha = 2/3$ 
solution of the subdiffusion equation. The procedure is neither 
very accurate nor of general use.    

Our aim here is to present a method to precisely measure $\alpha$ 
and $D_{\alpha}$. The method is described in detail in \cite{long}, 
here we give a brief account of it. For practical reasons, we choose 
for the experimental study a membrane system containing two vessels 
with a thin membrane in between which separates the initially 
homogeneous solute of the substance of interest from the pure solvent. 
A schematic view of the system is presented in Fig.~1. The membrane 
does not affect values of investigated parameters. Instead of the mean 
square displacement (\ref{a}), our method refers to the temporal 
evolution of the thickness $\delta$ of the so-called near-membrane 
layer which is defined as a distance from the membrane where the 
substance concentration $C(x,t)$ drops $\kappa$ times with respect 
to the membrane surface {\it i.e.}
\begin{equation}
\label{nml}
C(\delta ,t) = \kappa \: C( 0^+,t)\,,
\end{equation}
where $x=0$ is the position of a thin membrane and $\kappa$ is an 
arbitrary number smaller than unity; we used $\kappa=$ 0.12, 0.08 
and 0.05. 


\begin{figure}

\centerline{\epsfig{file=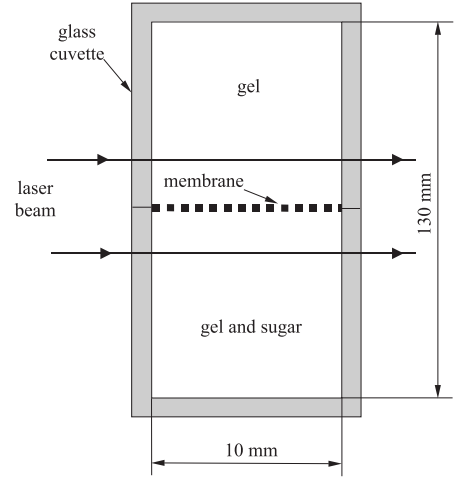,width=60mm}}

\vspace{2mm}

\caption{Schematic view of the membrane system under study.}

\end{figure}


In our previous paper \cite{dkwm}, we demonstrated that 
$\delta(t) = A  \sqrt{t}$ for the normal diffusion. Studying 
experimentally the diffusion of glucose and sucrose in a gel 
solvent, we show here that $\delta (t) = A \, t^{\gamma}$ with 
$\gamma < 0.5$. A gel is built of large and heavy molecules which
form a polymer network. Thus, the gel water solvent resembles a porous
material filled with water. Since a mobility of sugar molecules is 
highly limited in such a medium the subdiffusion is expected. 

For each measurement, we prepared two gel samples: the pure
gel - 1.5\% water solution of agarose and the same gel dripped
by the solute of glucose or sucrose. The concentration of both
sugars in the gel was fixed to be either
$0.1 \; [{\rm mol/dm^3}]$ or $0.07 \; [{\rm mol/dm^3}]$
but our results appear to be independent of the initial concentration.  
The two vessels of the membrane system were then filled with
the samples and the (slow) processes of the sugar transport across
the membrane started. Since the concentration gradient was in the 
vertical direction only, the diffusion is expected to be 
one-dimensional. We used an artificial membrane of the
thickness below 0.1 mm. The membrane was needed for two reasons.
It initially separated the homogeneous sugar solute in one vessel
from the pure gel in another one. It also precisely fixed the
geometry of the whole system.


\vspace{-1cm}

\begin{figure}

\hspace{2cm}
\centerline{\epsfig{file=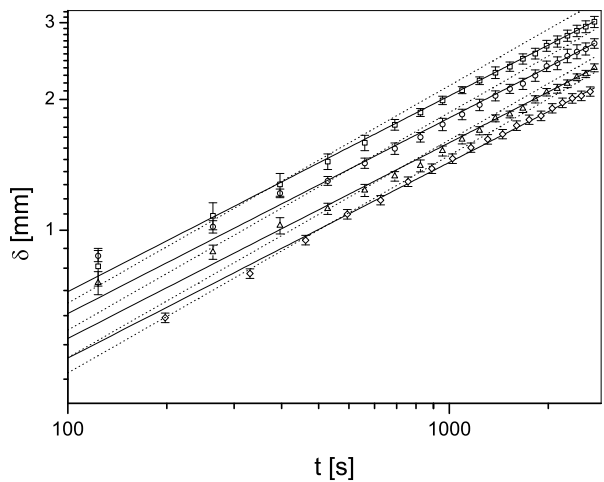,width=90mm}}

\caption{The experimentally measured thickness of the near-membrane layer
$\delta$ as a function of time $t$ for glucose with $\kappa=0.05$ ($\Box$),
$\kappa=0.08$ ($\circ$), $\kappa=0.12$ ($\triangle$), and for sucrose with
$\kappa=0.08$ ($\diamondsuit$). The solid lines represent the power 
function $A \: t^{0.45}$ while the dotted ones correspond to the function 
$A \sqrt{t}$.}

\end{figure}


The sugar concentration was measured by means of the laser 
interferometric method. The laser light was split into two beams. 
The first one went through the system parallelly to the 
membrane surface while the second (reference one) went directly to 
the light detecting system. The interferograms, which appear due 
to the interference of the two beams, are controlled by the 
refraction coefficient of the solute which is turn depends on the 
substance concentration. The analysis of the interferograms allows 
one to reconstruct the time-dependent concentration profiles of 
the substance transported in the system and to find the time 
evolution of the near-membrane layers which are of our main 
interest here. The experimental set-up is described in detail 
in \cite{kd}. It consists of the cuvette with membrane, the 
Mach-Zehnder interferometer including the He-Ne laser, TV-CCD 
camera, and the computerized data acquisition system.

When the sugar was diffusing across the membrane we were recording 
the concentration profiles in the vessel which initially contained 
pure gel. The examples of typical interferograms and extracted 
concentration profiles are presented in  \cite{kd}. The thickness 
of a near-membrane layer $\delta$ was calculated from the measured 
concentration profiles $C(x,t)$ according to the definition (\ref{nml}), 
and thus the thickness of near-membrane layer as a function of time 
was found. 

In Fig.~2 we present $\delta (t)$ for the glucose and sucrose of 
initial concentration $0.1 \; [{\rm mol/dm^3}]$. The analysis 
of errors, in particular those shown in Fig.~2, is described in 
\cite{long}. For the glucose we present $\delta (t)$ for three values 
of $\kappa=$ 0.12, 0.08 and 0.05 while for the sucrose $\kappa=0.08$. 
As seen, the time dependence of $\delta$ is well described by the 
power function $A\, t^\gamma$ with the common index $\gamma = 0.45$. 
The lines representing $\sim \! \sqrt{t}$ are also shown for comparison. 
It is evident that the measured index $\gamma$ is smaller than 0.5. 
There are some deviations of our data from $A\, t^{0.45}$ at 
$t < 300\:$s but our final theoretical formulas, in particular 
the power law behavior, hold in the long time approximation.

Fitting the experimental data shown in Fig.~2, we found the 
universal index $\gamma = 0.45 \pm 0.005$ and the parameter 
$A$ which depends on $\kappa$; for glucose $A = 0.091 \pm 0.004$ 
when $\kappa=0.05$, $A = 0.081 \pm 0.004$ when $\kappa=0.08$, 
and $A = 0.071 \pm 0.004$ when $\kappa=0.12$; for sucrose 
$A = 0.064 \pm 0.003$ when $\kappa=0.08$. In each case $\chi^2$ 
per degree of freedom was smaller than 1. 

The subdiffusion is described by the equation with fractional derivative 
\cite{mk,compte}
\be
\label{se}
\frac{\partial C(x,t)}{\partial t}=
D_{\alpha}\frac{\partial^{1-\alpha}}{\partial t^{1-\alpha}}
\frac{\partial^{2}C\left( x,t\right)}{\partial x^{2}}\,,
\ee
which for $\alpha < 1$ corresponds to an infinitely long average waiting 
time of the random walker - the physical situation in a gel solvent
resembling the porous medium. We solve Eq.~(\ref{se}) in the region 
$x>0$ with the initial condition $C(x,0) = C_0$ for $x<0$ and 
$C(x,0) = 0$ for $x>0$. In fact, we solve Eq.~(\ref{se}) for the 
Green's function $G(x,t;x_0)$ satisfying the initial condition 
$G(x,t=0;x_0)=\delta(x-x_0)$, and then, $C(x,t)$ is calculated 
using the formula
\be
\label{int}
C(x,t) =\int G(x,t;x_0) \: C(x_0,0) \: dx_0\,.
\ee
To find $G(x,t;x_0)$ we use the relation \cite{long}
\be
\label{gint}
G(x,t;x_0) =\int_0^{t} dt'\:  J(0^+,t';x_0) \:
G_{\rm ref}( x,t-t';0^+) \,,
\ee
where $x > 0$ while $x_0 < 0$; $J(x,t;x_0)$ is the flux associated
with $G(x,t;x_0)$ which for $x=0$  gives the flow across the membrane;
$G_{\rm ref}(x,t;x_0)$ is the Green's function for the half-space system
with $x > 0$ and the fully reflecting wall at $x=0$. 

Using Eqs.~(\ref{int},\ref{gint}), $C(x,t)$ can be written as
\be
\label{prof}
C(x,t) =\int_0^t dt' \, W(t') \:
G_{\rm ref}(x,t-t';0^+) \,,
\ee
where the function $W(t)$, which equals
$$
W(t) =\int_{- \infty}^0 dx_0 \,  J(0^+,t;x_0) \:
C( x_0,0) \,,
$$
depends on the initial and boundary conditions.

Since the subdiffusion equation is of the second order with respect 
to $x$, it requires two boundary conditions at the membrane. The first 
one assumes the continuity of the flux $J$, given by the generalized 
Fick's law \cite{z}, which flows through the membrane {\it i.e.} 
$J(0^-,t) =J(0^+,t)$. However, there is no obvious choice of the 
second boundary condition. Therefore, we assume that the missing 
condition is given by a linear combination of 
concentrations and flux {\it i.e.}
\be
\label{general-boundary}
b_1 C(0^-,t) + b_2 C(0^+,t) + b_3 J(0,t) = 0 \,.
\ee
Two boundary conditions 
\be
\label{nbc}
C(0^+,t)=\frac{1-\sigma}{1+\sigma} \: C(0^{-},t)\,,
\ee
and 
\be
\label{rbc}
J(0,t) =\lambda \left( C(0^-,t) -C(0^+,t)\right)\,,
\ee
discussed in \cite{dkwm,kpr} and  \cite{kmapp}, respectively,
are of the general form (\ref{general-boundary}). The parameters 
$\sigma$ and  $\lambda$ control the membrane permeability 
\cite{dkwm,kpr,kmapp}. The adopted initial condition combined 
with Eq.~(\ref{general-boundary}) provide
\ba
\label{W-exp}
W (t) &=& C_0 \frac{b_1 \sqrt{D_\alpha}}{b_1 - b_2}
\frac{1}{t^{1-\alpha/2}} 
\\ \nonumber 
&\times& \sum_{k=0}^{\infty}
\frac{d^k}{\Gamma(\alpha/2 - k(1-\alpha/2))}
\frac{1}{t^{k(1-\alpha/2)}}\,,
\ea
where $d \equiv b_3 \sqrt{D_\alpha}/(b_1 - b_2)$. The Green's 
function $G_{\rm ref}$, which enters Eq.~(\ref{prof}),
can be easily obtained by means of the method of images \cite{mk} 
as $G_{\rm ref}(x,t;x_0) =G_0(x,t;x_0) +G_0(-x,t;x_0)$ with 
the known Green's function $G_0$ for the homogeneous system \cite{mk}. 
Having the explicit functions $W$ and $G_{\rm ref}$, we write 
down, using Eq.~(\ref{prof}), the concentration profile as
\ba
\label{profil-final}
C(x,t) &=& \int_0^t dt'\: W(t-t') 
\\ \nonumber
&\times&
\frac{2}{\alpha x} \:
H_{1\;1}^{1\;0} \bigg(\Big(\frac{x}{\sqrt{D_{\alpha}{t'}^{\alpha}}}
\Big)^{\frac{2}{\alpha}} \bigg|
\begin{array}{cc}
1 & 1 \\
1 & \frac{2}{\alpha}
\end{array}
\bigg)\,,
\ea
where $H$ denotes the Fox function. 

We first consider the long time approximation of the formula
(\ref{profil-final}) which corresponds to the small $s$ limit 
of the Laplace transform $L\left\{ f( t)\right\} \equiv 
\int_{0}^{\infty}f(t)e^{-st}$. Taking into account only the leading 
contribution in the small $s$ limit, Eq.~(\ref{profil-final}) gets 
the form
\be
\label{profil-long}
C(x,t) = \frac{2 C_0 b_1}{(b_1 -b_2)\alpha} \:
H_{1\;1}^{1\;0} \bigg(\Big(\frac{x}{\sqrt{D_{\alpha}t^{\alpha}}}
\Big)^{\frac{2}{\alpha}} \bigg|
\begin{array}{cc}
1 & 1 \\
0 & \frac{2}{\alpha}
\end{array}
\bigg)\,.
\ee
The solution (\ref{profil-long}) can be also obtained directly from
Eq.~(\ref{profil-final}), taking into account only the $k=0$ term in
the expansion (\ref{W-exp}).

The series (\ref{W-exp}) can be approximated by the first term if
$d \ll t^{1-\alpha/2}$. When the boundary condition is of the form 
(\ref{nbc}), the condition is trivially satisfied for any $t$ as 
$b_3=d=0$ in this case. For the boundary condition (\ref{rbc}), 
we have $\lambda = b_1/b_3 = - b_2/b_3$, and the long time approximation 
holds if
\be
\label{long-cond2}
\bigg(\frac{\sqrt{D_\alpha}}{2 \lambda}
\bigg)^{\frac{1}{1-\alpha/2}} \ll t \,.
\ee
For the membranes used in our experiments $\lambda$ is of order 
$10^{-2} \; {\rm [mm/s]}$ and assuming that we deal with the normal 
diffusion $D$ is roughly $10^{-5} \; {\rm [mm^2/s]}$. Thus, the l.h.s.
of Eq.~(\ref{long-cond2}) is estimated as 2~s. Since 10~s is the time
step of our measurements which extend to 2500~s, the condition
(\ref{long-cond2}) is fulfilled. We have also checked the condition
(\ref{long-cond2}) {\it a posteriori}, using the values of $\alpha$
and $D_\alpha$ obtained by means of our method. The l.h.s. of
Eq.~(\ref{long-cond2}) is again about 2~s.

Let us now discuss the temporal evolution of near-membrane layers in the
long time approximation. Substituting the solution (\ref{profil-long})
into Eq.~(\ref{nml}), we get the equation which simplifies to
\be
\label{nml-eq}
H_{1\;1}^{1\;0} \bigg(\Big(\frac{\delta}{\sqrt{D_{\alpha}t^{\alpha}}}
\Big)^{\frac{2}{\alpha}} \bigg|
\begin{array}{cc}
1 & 1 \\
0 & \frac{2}{\alpha}
\end{array} \bigg)
= \frac{\kappa \alpha}{2}\,.
\ee
One observes that Eq.~(\ref{nml-eq}) is solved by
\be
\label{nml-sol}
\delta(t) = A(\alpha,D_{\alpha},\kappa) \: t^{\alpha/2} \,.
\ee
The near-membrane layer (\ref{nml-sol}) does not depend on the 
parameters $b_1$ and $b_2$ while the coefficient $A$ can be 
recalculated into the diffusion constant $D_\alpha$ as
\be
\label{sc}
D_{\alpha}=\frac{A^2}{\left[ \big(H_{1\;1}^{1\;0}\big)^{-1}
\Big( \frac{\alpha\kappa}{2} \Big|
\begin{array}{cc}
1 & 1 \\
0 & \frac{2}{\alpha}
\end{array}
\Big) \right]^{\alpha}}\,.
\ee

We have also studied the near-membrane layers beyond the long time
approximation using the boundary conditions (\ref{nbc}) and 
(\ref{rbc}). The condition (\ref{nbc}) allows for the 
analytic treatment of Eq.~(\ref{se}) and the solution is of the 
form (\ref{profil-long}) with $2b_1/(b_1 - b_2)$ replaced by 
$1 - \sigma$. Thus, the formulas derived in the long time 
approximation are exact for Eq~(\ref{nbc}). When Eq.~(\ref{rbc}) 
is used as the boundary condition, the solution of subdiffusion 
equation (\ref{se}) for $x>0$ is
\ba
\label{cr}
C(x,t) &=& \frac{C_0}{\alpha} \:
\sum_{n=0}^{\infty}\bigg[ -\frac{x}{2\lambda}
\Big( \frac{\sqrt{D_{\alpha}}}{x}\Big)^\frac{2}{\alpha}\bigg]^n \:
\\ \nonumber
&\times&
H_{1\;1}^{1\;0}
\bigg(\Big(\frac{x^{2}}{D_{\alpha}t^{\alpha}}\Big)^{\frac{1}{\alpha}}
\bigg|
\begin{array}{cc}
1 & 1 \\
n\left(\frac{2}{\alpha}-1\right) & \frac{2}{\alpha}
\end{array}
\bigg)\,.
\ea
The solutions (\ref{profil-long}) and (\ref{cr}), which for normal 
diffusion have been discussed in \cite{kpr,kmapp}, qualitatively 
differ from each other but the differences are evident 
only for times which are significantly longer than those
studied here.

Since the formula (\ref{cr}) is analytically intractable we have found
the time evolution of near-membrane layer numerically. As discussed in 
detail in \cite{long}, we have not found any difference between the
near-membrane layer obtained for the concentration profile with the 
boundary condition (\ref{nbc}) and with the boundary condition (\ref{rbc}). 

Fitting the experimental $\delta (t)$ by the function
$A\,t^\gamma$, we have found the index 
$\alpha = 2\gamma = 0.90 \pm 0.01$. It does not much differ
from unity but it signals subdiffusion due to the small error
\cite{long}. With the numerical values of inverse Fox functions, 
we recalculate the coefficient $A$ into $D_\alpha$ by means of 
the relation (\ref{sc}). Thus, we get 
$D_{0.90}=(9.8 \pm 1.0)\times 10^{-4}\; [{\rm mm^2/s^{0.90}}]$
for glucose and 
$D_{0.90}=(6.3 \pm 0.9)\times 10^{-4}\; [{\rm mm^2/s^{0.90}}]$
for sucrose. 


\vspace{-1cm}

\begin{figure}

\hspace{2cm}
\centerline{\epsfig{file=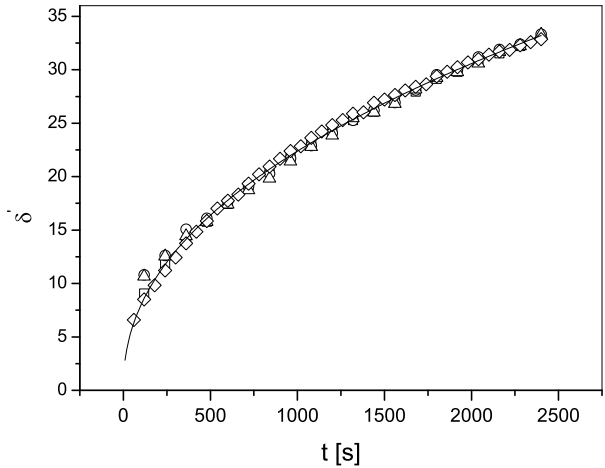,width=90mm}}

\caption{The experimentally measured $\delta$ divided by the coefficient
$A$ from Eq.~(\protect\ref{nml-sol}). The symbols are assigned as in
Fig.~2 and the line represents the function $t^{0.45}$. For clarity 
of the plot the error bars are not shown.} 

\end{figure}


To be sure that Eq.~(\ref{nml-sol}), which is used to evaluate $D_\alpha$,
properly describes the experimental $\delta (t)$, we have checked the 
scaling of $\delta (t)$ suggested by Eq.~(\ref{nml-sol}). In Fig.~3 we 
plot the rescaled near-membrane layer $\delta'(t) = \delta(t)/ A$, with 
$A$ from Eq.~(\ref{nml-sol}), for all values of $\kappa$, for glucose 
and for sucrose. The experimental points are represented as in Fig.~2. 
As seen, our experimental data are very well described by the 
function $t^{0.45}$.

Our method to determine the parameters of subdiffusion relies 
on the near-membrane layers. One may ask why $\alpha$ and 
$D_{\alpha}$ are not extracted directly form the concentration 
profiles which are measured. There are three reasons to choose 
the near-membrane layers: experimental, theoretical and practical:
1) Measurement of $\delta$ does not suffer from the 
sizable ($\sim$ 10-15\%) systematic error of absolute 
normalization of $C$, as only the relative concentration matters 
for $\delta$. 2) Computed concentration profiles depend on the adopted 
boundary condition at a membrane while the condition is not well 
established even for the normal diffusion. The near-membrane 
layer appears to be free of this dependence. 3) When $C$ is fitted 
by a solution of the subdiffusion equation, there are three free 
parameters: $\alpha$, $D_{\alpha}$ and the parameter characterizing 
the membrane permeability. Because these fit parameters are correlated 
with each other it is very difficult to get their unique values. 
When $\delta$ is studied the membrane parameter drops out entirely, 
$\alpha$ is controlled by the time dependence of $\delta (t)$ while 
$D_{\alpha}$ is provided by the coefficient $A$. 

The membrane plays only an auxiliary role in our method to measure
the subdiffusion parameters but the transport in membrane systems is 
of interest in several fields of technology \cite{Rau89}, where the 
membranes are used as filters, and biophysics \cite{Tho76}, where the 
membrane transport plays a crucial role in the cell physiology. 
The diffusion in a membrane system is also interesting by itself as 
a nontrivial stochastic problem, see {\it e.g.} \cite{kpr}. Thus, 
our study of the subdiffusion in a membrane system, which to our best 
knowledge has not been investigated by other authors, opens up a new 
field of interdisciplinary research. 

\vspace{-0.5cm}


\end{narrowtext}

\begin{thebibliography}{50}


\vspace{-1.5cm}

\bibitem{mk} R. Metzler and J. Klafter, Phys. Rep.  {\bf 339}, 1 (2000).

\bibitem{bg} J.P. Bouchaud and A. Georgies, Phys. Rep. {\bf 195} (1990), 127.

\bibitem{bmk} E. Barkai, R. Metzler, and J. Klafter, Phys. Rev. E {\bf 61}, 
132 (2000); E. Barkai, Phys. Rev. E {\bf 63}, 046118 (2001);
S. Lim and S.V. Muniandy, Phys. Rev. E {\bf 66}, 021114 (2002).

\bibitem{km} A. Klemm, R. Metzler, and R. Kimmich, Phys. Rev. E {\bf 65}, 
021112 (2002).

\bibitem{long} T. Koszto{\l}owicz, K. Dworecki, and
St. Mr\'owczy\' nski, [arXiv:cond-mat/0309072], submitted
to Phys. Rev. {\bf E}.

\bibitem{dkwm} K. Dworecki, T. Koszto{\l}owicz, S. W\c{a}sik, and 
St. Mr\'owczy\' nski, Eur. J. Phys. E {\bf 3}, 389 (2000).

\bibitem{kd} K. Dworecki, J. Biol. Phys. {\bf 21}, 37 (1995).

\bibitem{compte} A. Compte, Phys. Rev. E {\bf 53}, 4191 (1996).

\bibitem{z} D.H. Zanette, Physica A {\bf 252}, 159 (1998).

\bibitem{kpr} T. Koszto{\l}owicz, Phys. Rev. E {\bf 54}, 3639 (1996);
J. Phys. A {\bf 31}, 1943 (1998); Physica A {\bf 248}, 44 (1998).

\bibitem{kmapp} T. Koszto{\l}owicz and St. Mr\'owczy\'nski, 
Acta Phys. Pol. B {\bf 32}, 217 (2001);
T. Koszto{\l}owicz, Physica A {\bf 298}, 285 (2001).

\bibitem{Rau89} R. Rautenbach and R. Albert, {\it Membrane Processes},
(Wiley, Chichester, 1989).

\bibitem{Tho76} J.H.M. Thornley, {\it Mathematical Models in Plant Physiology},
(Academic Press, London, 1976).


\end{thebibliography}
\end{document}